\documentclass[epj]{webofc}
\usepackage[varg]{txfonts}   

\usepackage{lineno}
\doi{ 10.1051/epjconf/201713603013}
\begin{document}
%
%
\title{LATTES: a new gamma-ray detector concept for South America}
%
%

\author{P. Assis\inst{1} \and
           U. Barres de Almeida\inst{2} \and
           A. Blanco\inst{3} \and
           R. Concei\c{c}\~{a}o\inst{1}\fnsep\thanks{\email{ruben@lip.pt  (presenter)}} \and
            B. D'Ettoree Piazzoli\inst{4} \and
            A. De Angelis\inst{5,6,1} \and
            M. Doro\inst{7,5} \and
            P. Fonte\inst{3,8} \and
            L. Lopes\inst{3} \and
            G. Matthiae\inst{7} \and
            M. Pimenta\inst{1} \and
            R. Shellard\inst{2} \and
             B. Tom\'e\inst{1}
}

\institute{LIP Lisboa and IST Lisboa, Portugal
\and
CBPF, Rio de Janeiro, Brazil
\and
LIP Coimbra and University of Coimbra, Portugal
\and
Universit\`a  di Napoli ``Federico II'' and INFN Roma Tor Vergata, Italy
\and
INFN Padova, Italy
\and
Universit\`a di Udine, ItalyUniversit\`a di Padova, Italy
\and
INFN and Universit\`a di Roma Tor Vergata, Roma, Italy
\and
Coimbra Polytechnic - ISEC, Coimbra, Portugal
          }

\abstract{%
Currently the detection of Very High Energy gamma-rays for astrophysics rely on the measurement of the Extensive Air Showers (EAS) either using Cherenkov detectors or EAS arrays with larger field of views but also larger energy thresholds. In this talk we present a novel hybrid detector concept for a EAS array with an improved sensitivity in the lower energies ($\sim 100\,$GeV). We discuss its main features, capabilities and present preliminary results on its expected perfomances and sensitivities.This wide field of view experiment is planned to be installed at high altitude in South America making it a complementary project to the planned Cherenkov telescope experiments and a powerful tool to trigger further observations of variable sources and to detect transients phenomena.}
\maketitle
\section{Introduction}
\label{intro}
\par The study of high-energy and very-high-energy gamma-rays is very important to probe extreme phenomena that takes place in the Universe. Moreover, being neutral, this radiation can pin-point to their emission source, as they are not deflect by the surrounding magnetic fields.
\par The detection of gamma-rays at lower energies (below $\sim 100\,$GeV) can be done using instruments placed in artificial satellites, for instance Fermi. However, as the gamma-ray energy increases, its flux at Earth becomes increasingly smaller, to the point where the available collection areas aboard satellites are not enough to study them. Fortunately, the interaction of gamma-rays with the Earth atmosphere produces Extensive Air Shower (EAS) whose secondaries can be sampled by detector arrays, or one could collect the Cherenkov light produced by the secondaries using Imaging Array Cherenkov Telescopes (IACTs). These ground based detectors attain different advantages/disadvantages with respect to each other: IACTs have a lower energy threshold,  and have better angular and energy resolution, as they can image the shower development; on the other hand EAS array have significantly wider field-of-views.
\par
\begin{figure}[ht]
\centering
\sidecaption
\includegraphics[scale=0.4]{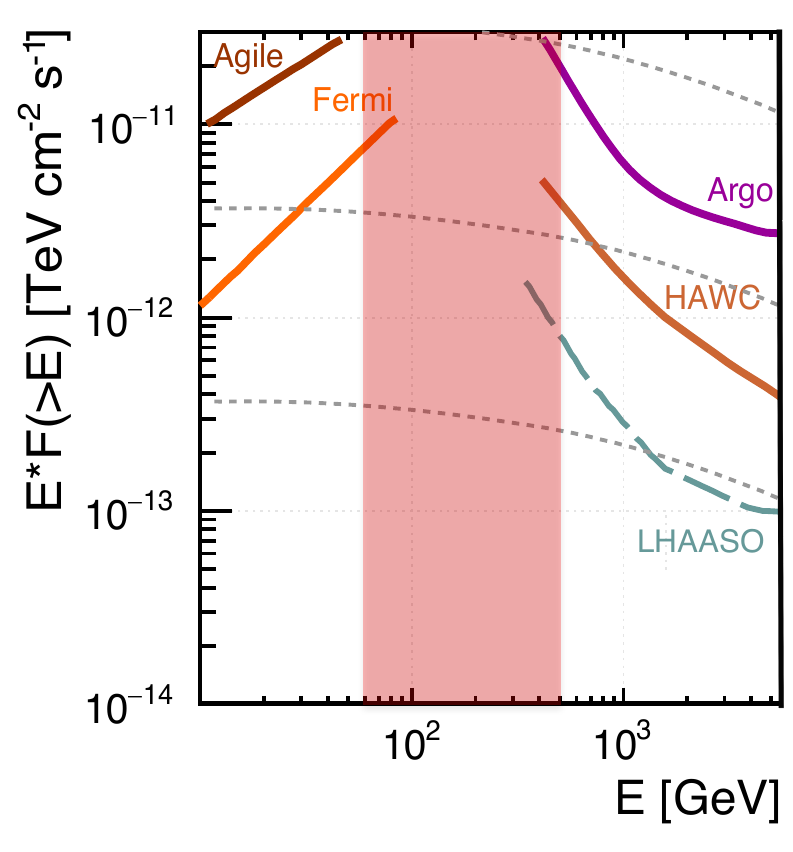}
\caption{Integrated sensitivity of several gamma-ray experiments with a wide field-of-view as a function of the primary energy. (Taken from~\cite{pimenta}.)}
\label{fig:Espectrum}
\end{figure}

 In figure \ref{fig:Espectrum} it is shown the sensitivities of current and future gamma-ray experiments with wide field-of-views. Two things become evident: there is no wide field-of-view experiment covering the Southern hemisphere sky; there is a gap in the $100\,$GeV region. 
Such wide FoV experiment with a low energy threshold and a large duty cycle would be fully complementary to the powerful narrow-FoV Cherenkov Telescope Array (CTA) as it would be able not only to issue alerts of transient phenomena but would also enable long term observations of variable sources and help on the search for emissions from extended regions, such as the Fermi bubbles or dark matter annihilations from the centre of our galaxy. Hence, we propose a novel hybrid detector to be installed at $\sim 5200 \,$m a.s.l which ensures an improved sensitivity to the $100\,$ GeV energy region. 
\par This manuscript is organised as follows: in section \ref{sec:detector} we describe the detector and the layout of the experiment. In section \ref{sec:exp} we discussed the performance of such detector and in section \ref{sec:sensitivity} we present the achieved sensitivities. We end with final remarks,
\vspace{-1cm}

\section{Detector description}
\label{sec:detector}
\par
\begin{figure}[ht]
\centering
\sidecaption
\includegraphics[scale=0.20]{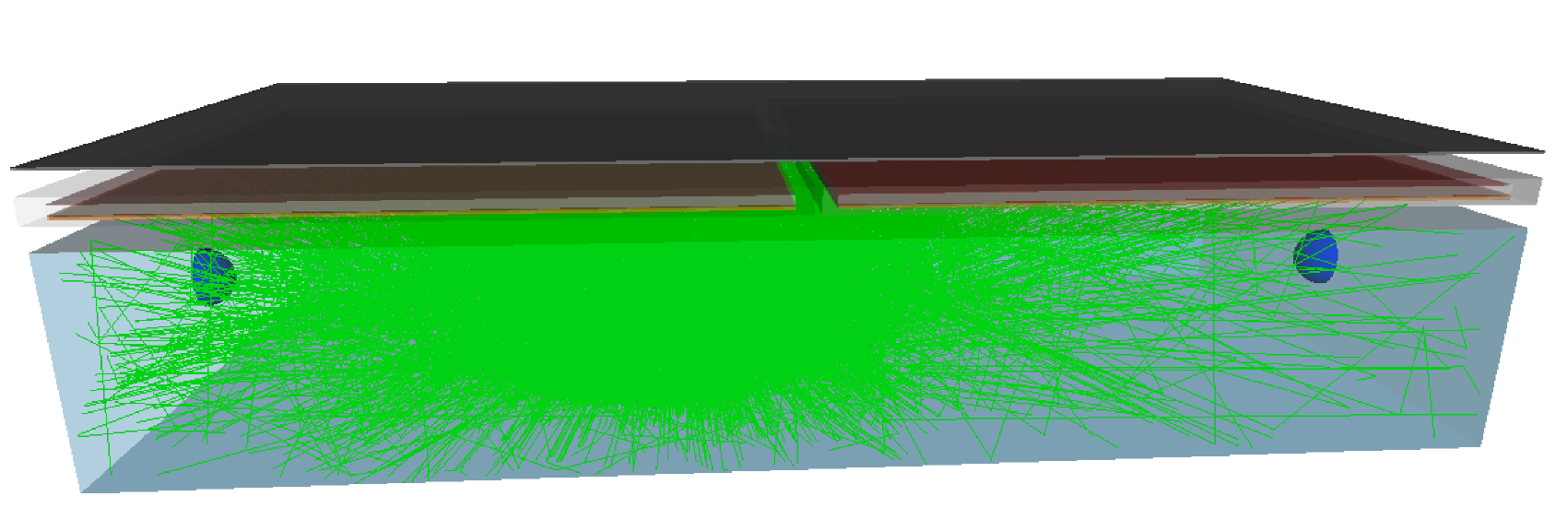}
\caption{Basic detector station with 1 WCD covered with RPCs and a thin lead plate. The green lines show the tracks of Cherenkov photons produced by charged particles in the WCD. The blue semi-spheres inside the WCD are the two PMTs.}
\label{fig:station}
\end{figure}

 In order to surpass the limitations of the previous EAS arrays experiments, and be able to lower the energy threshold while maintaining a reasonable energy and angular resolution, we propose to build a dense array with an area of $10\,000\,{\rm m^2}$ constituted by modular hybrid detectors. Each station is composed by two low-cost Resistive Plate Chambers (RPC) on top of a Water Cherenkov Detector (WCD), as shown in figure~\ref{fig:station}. Each RPC has 16 charge collecting pads covering a total area of $1.5\times.1.5\,{\rm m^2}$. The WCD has a rectangular structure with dimensions $3 \times 1.5\times0.5\,{\rm m^3}$. The signals are read by two photomultipliers (PMTs) at both ends of the smallest vertical face of the WCD. On the top of the RPCs there is a thin lead plate ($5.6\,$mm) to convert secondary photons. The conversion of the photons is important to improve the geometrical reconstruction as photons have a stronger correlation with the primary direction with respect to secondary shower electrons. The success of such hybrid detector concept lies on the fact that the RPCs contribute with its high segmentation and time resolution while the WCD provides a calorimetric measure of the shower secondary particles allowing to lower the energy threshold. Moreover, with this detector concept, it is possible to trigger in the WCD which allows the RPCs to operate at a low threshold while minimising several sources of noise (detector, electronics, environment).

\section{Experiment performance}
\label{sec:exp}
\par The performance of this detector has been assessed using an end-to-end realistic Monte Carlo simulation. The EAS have been simulated using CORSIKA (COsmic Ray Simulations for KAscade) and the detector response was treated by a Geant4 dedicated simulation.
\par We generated 10000 CORSIKA simulations for gammas and protons between $10\,$GeV and $5\,$TeV. To save computational times, the simulations were generated using a power law differential energy spectrum with an index $-1$, and afterwards were  weighted for the corresponding particle fluxes. The zenith angle for gammas was fixed to $10^{\circ}$, while for protons the range was between 5 and 15 degrees. The detector was assumed to be placed at an altitude of $5200\,$m a.s.l.
\par
\begin{figure}[ht]
\centering
\includegraphics[scale=0.35]{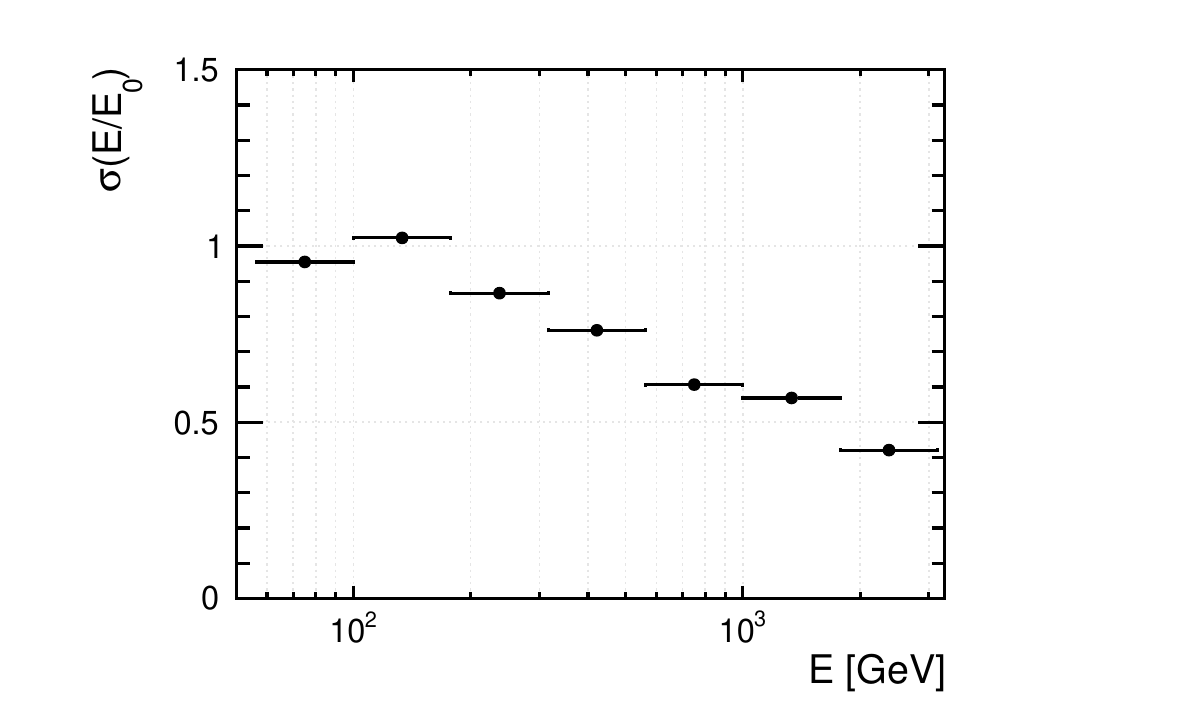}
\includegraphics[scale=0.35]{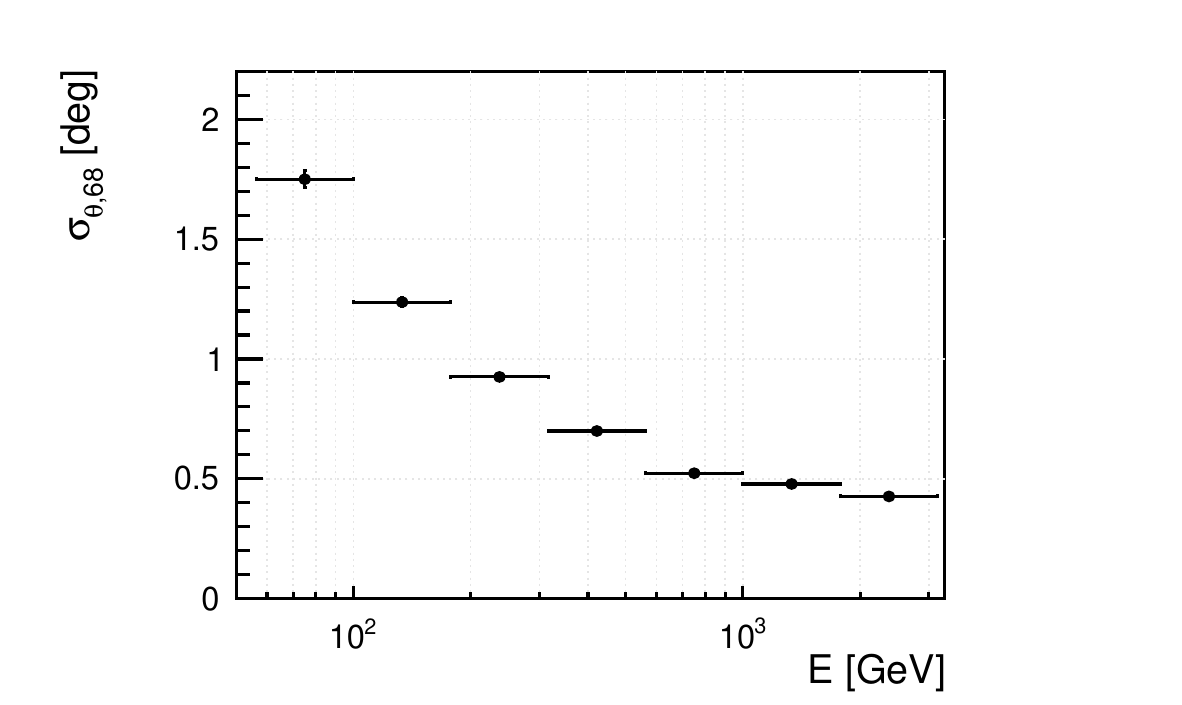}
\caption{ (Left) Reconstructed energy resolution as a function of the reconstructed energy for proton-initiated showers. (Right) Angular resolution for gamma-ray primaries with a zenith angle of $10^{\circ}$ as a function of the reconstruction energy.}
\label{fig:angular}
\end{figure}

 To evaluate the effective area at the trigger level, we required that at least 3 stations have detected a signal. The trigger condition for the station requires at least 5 photonelectrons in each PMT. The effective area for this array, has been computed using simulations. We have found that for gamma primaries with an energy of 100 GeV we still have around $10^{3}\,$m$^2$, even considering selection quality cuts, which will be described below. The energy estimation has been done using the total signal  ($S_{tot}$) recorded in all WCD stations for each event. A calibration curve was derived and used to get the reconstructed energy ($E_{rec}$) for each event from $S_{tot}$. From this curve, it is possible to evaluate the energy resolution that one could achieve with such detector as a function of $E_{rec}$. This is shown in figure \ref{fig:angular} (left) where it is possible to see that the energy resolution improves as the shower energy increases and, at the lowest energies, one still has a reasonable energy resolution of $100\%$, being this mostly dominated by shower-to-shower fluctuations. 
\par The geometric reconstruction was done taking advantage of the RPC segmentation and fast timing (it was used in the simulations a time resolution of $1\,$ns). The position and time of the recorded hits in the RPC were fitted to a shower front plane model in order to the reconstruct the primary direction. The quality of the reconstruction can be improved applying a cut on the number of active RPCs' pads in the event: it was required that the event has at least 10 hits. The reconstructed angle was compared to the simulated one, and we calculate the $68\%$ containment angle, $\sigma_{\theta,68}$. The result is shown in figure \ref{fig:angular} (right) where it can be seen that, at energies around $100\,$GeV, a reasonable resolution can be achieved, better that $2^{\circ}$.

\section{Sensitivity}
\label{sec:sensitivity}
\par In order to compute this detector sensitivity to steady sources, one needs to know, apart from the efective area, the energy and angular resolution, the discrimination capabilities between gamma and hadrons. Although we strongly believe that this hybrid detector could combine strategies explored in previous experiments such, as HAWC~\cite{HAWCgh} and ARGO\cite{ARGOgh}, the complexity of such required study is out of the scope of this manuscript. Therefore, conservatively, we assumed no background rejection below $300\,$GeV. As in this manuscript we aim to focus on the lowest energies, above $300\,$GeV we took HAWC gamma/hadron capabilities as an ansatz for the highest energies~\cite{HAWCgh}. This should, of course, be carefully studied in a future work.
\par
\begin{figure}[ht]
\centering
\sidecaption
\includegraphics[scale=0.45]{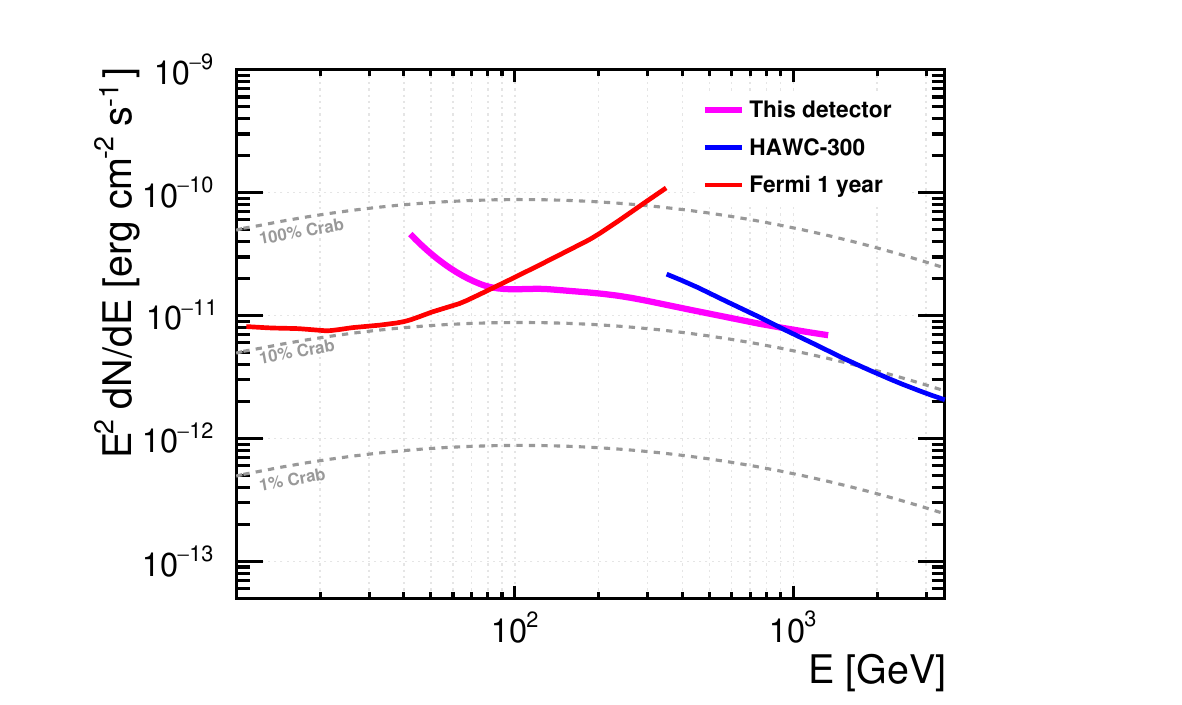}
\caption{Differential sensitivity for steady sources after one year of effective time. See legend for details.}
\label{fig:sensitivity}
\end{figure}

 In figure \ref{fig:sensitivity} it is shown the differential sensitivity of this detector to study sources. We compute the sensitivity as the flux of a source giving $N_{excess}/\sqrt{N_{bkg}} = 5$ after 1 year of effective observation. It was assumed that the source is visible one fourth of the time, which is roughly the time that the galactic centre is visible in the Southern tropic. The obtained results are compared with the 1 year sensitivities of FERMI and HAWC. One can clearly see that this detector would be able to cover the gap between the two of most sensitive experiments in this energy range.

\section{Final remarks}
\label{sec:final}
\par We have presented a novel hybrid detector able to extend of previous experiments down to the region of $100\,$GeV (more information can be found in \cite{our}). This modular compact and low cost detector has given encouraging results, but its capabilities are far from explored, in particular, in what respects gamma-hadron discrimination. With the advent of the Cherenkov Telescope Array, this experiment would be a complementary project as it could provide not only triggers to transient phenomena as it could do long term observations of variable sources.

\textbf{Acknowledgments}
R. Concei\c{c}\~ao acknowledges the financial support given by FCT-Portugal \mbox{SFRH/BPD/73270/2010}.

%
 \bibliography{biblio.bib}
%

\end{document}